\providecommand{\main}{.}
\documentclass[12pt, draftclsnofoot, journal, onecolumn,romanappendices]{IEEEtran}
	\usepackage[T1]{fontenc}		
	\usepackage[utf8]{inputenc}	
	\usepackage[english]{babel}	

	\usepackage{url}
	\usepackage{ifthen}

	\usepackage{amsthm,amssymb}
	\usepackage[cmex10]{amsmath} 
	\usepackage{bm}
	\usepackage{nicefrac}
	\usepackage{dsfont} 
	\usepackage[subnum]{cases}  

	\usepackage{subfiles} 

	\usepackage{placeins} 	
	\usepackage{caption}
	\usepackage{subcaption}
	\usepackage{tikz}
		\usetikzlibrary{arrows,positioning,decorations.pathreplacing,calligraphy}
		\usetikzlibrary{fit,shapes}
		\usetikzlibrary{calc}

	\usepackage{pgfplots}
		\usepgfplotslibrary{fillbetween}
	\usepackage[percent]{overpic}
	\usepackage{pict2e} 
	\usepackage[most]{tcolorbox}
	
	\usepackage{multirow}
	\usepackage{colortbl}
	\usepackage{longtable} 	
	\usepackage{array}
	\usepackage{arydshln}  
	\usepackage{diagbox}
	
	\usepackage[final]{microtype} 
	\usepackage{tablefootnote} 
	\usepackage{enumitem}
	
	\usepackage{color}
	\definecolor{gray}{rgb}{0.6,0.6,0.6}

	\definecolor{blued}{rgb}{0.0,0.5,1}	
	\definecolor{greened}{rgb}{0.28,.71,.46}	
	\usepackage{hyperref}	
	\usepackage{cite}

	\newtheorem{theorem}{Theorem}
	\newtheorem{lemma}{Lemma}
	\newtheorem{proposition}{Proposition}
	\newtheorem{definition}{Definition}
	\newtheorem{remark}{Remark}

	\allowdisplaybreaks  
	\makeatletter 
	\newcommand{\leqnomode}{\tagsleft@true}
	\newcommand{\reqnomode}{\tagsleft@false}
	\makeatother

	\newtoggle{draft}
	\toggletrue{draft}

\definecolor{mycolor1}{rgb}{0.00000,0.44700,0.74100}%
\definecolor{mycolor2}{rgb}{0.85000,0.32500,0.09800}%
\definecolor{mycolor3}{rgb}{0.92900,0.69400,0.12500}%
\definecolor{mycolor4}{rgb}{0.49400,0.18400,0.55600}%
\definecolor{mycolor5}{rgb}{0.300000,0.64314,0.00000}%
\colorlet{mygreen}{green!60!black}%
\colorlet{myblue}{mycolor1!80!black}%

\newcommand{\Ybar}{\ensuremath{\bar{Y}}}
\newcommand{\bYbar}{\ensuremath{\bar{\bY}}}
\newcommand{\Xbar}{\ensuremath{\bar{X}}}
\newcommand{\bXbar}{\ensuremath{\bar{\bX}}}

\newcommand{\nt}{^{[n]}}

	\definecolor{mycolor1}{rgb}{0.00000,0.44700,0.74100}%
	\definecolor{mycolor2}{rgb}{0.85000,0.32500,0.09800}%
	\definecolor{mycolor3}{rgb}{0.92900,0.69400,0.12500}%
	\definecolor{mycolor4}{rgb}{0.49400,0.18400,0.55600}%
	\definecolor{mycolor5}{rgb}{0.300000,0.64314,0.00000}%
	
\newcommand\itemEq[1][]{%
  \ifx\relax#1\relax  \item \else \item[#1] \fi
  \abovedisplayskip=0pt\abovedisplayshortskip=0pt~\vspace*{-\baselineskip}}
	
\newcommand{\myoverset}[2]{\ensuremath{\overset{\mathclap{#1}}{#2}}}

\newcommand{\itb}{\begin{itemize}}
\newcommand{\ite}{\end{itemize}}
\newcommand{\enb}{\begin{enumerate}}
\newcommand{\ene}{\end{enumerate}}
\newcommand{\eqm}[1]{\begin{align}#1\end{align}}
\newcommand{\eqmo}[1]{\begin{equation}\begin{aligned}#1\end{aligned}\end{equation}}

\newcommand{\limpf}{\lim_{P\rightarrow\infty}}
	
\newcommand{\Pb}{{\bar{P}}}

\DeclareMathAlphabet{\mathbit}{OML}{cmr}{bx}{it}
\DeclareMathAlphabet{\mathsf}{OT1}{cmss}{m}{n}
\DeclareMathAlphabet{\mathTXf}{OT1}{cmss}{bx}{it}

\DeclareMathOperator{\DoF}{DoF}

\DeclareMathOperator{\RX}{RX}
\DeclareMathOperator{\TX}{TX}

\newcommand{\Zb}{{{\mathbb{Z}}}}

\newcommand{\Cb}{{{\mathbb{C}}}}
\newcommand{\Cc}{{{\mathcal{C}}}}
\newcommand{\Dc}{{{\mathcal{D}}}}

\newcommand{\Gc}{{{\mathcal{G}}}}
\newcommand{\Hc}{{\mathcal{H}}}
\newcommand{\Ic}{{{\mathcal{I}}}}

\newcommand{\Sc}{{{\mathcal{S}}}}

\newcommand{\bA}{\mathbf{A}}

\newcommand{\bH}{\mathbf{H}}
\newcommand{\bI}{\mathbf{I}}

\newcommand{\bL}{\mathbf{L}}

\newcommand{\bN}{\mathbf{N}}

\newcommand{\bR}{\mathbf{R}}

\newcommand{\bX}{\mathbf{X}}
\newcommand{\bY}{\mathbf{Y}}

\newcommand{\trans}{{\text{T}}}

\usepackage{cleveref} 
\usepackage[overload]{empheq}

\newcommand{\PD}{\mathrm{PD}}
\newcommand{\bTX}{\mathrm{\textbf{TX}}}

\newcommand{\minA}{\ensuremath{m^{M-k}_{N_2}}}

\togglefalse{draft}

\begin{document}

		\title{DoF Region of the Decentralized MIMO Broadcast Channel---How many informed antennas do we need?} 
 \author{%
   \IEEEauthorblockN{ Antonio Bazco-Nogueras\IEEEauthorrefmark{1}\IEEEauthorrefmark{3},
                     Arash  Davoodi\IEEEauthorrefmark{2},
                     Paul de Kerret\IEEEauthorrefmark{1},
                     David Gesbert\IEEEauthorrefmark{1},
                     Nicolas Gresset\IEEEauthorrefmark{3}
										 and
                     Syed Jafar\IEEEauthorrefmark{4}										
										}
										
	\normalsize
									
   \IEEEauthorblockA{\IEEEauthorrefmark{1}%
                     Communication Systems Department, EURECOM,
                     Sophia Antipolis, France, \\
                     Email: \{bazco, dekerret, gesbert\}@eurecom.fr}
										
   \IEEEauthorblockA{\IEEEauthorrefmark{2}%
										 Department of Computational Biology, 
										 Carnegie Mellon University,
										 Pittsburgh, PA, 
                     USA,\\
                     Email: agholami@andrew.cmu.edu}
											
   \IEEEauthorblockA{\IEEEauthorrefmark{3}%
                     Mitsubishi Electric R\&D Centre Europe (MERCE), 
										35700 Rennes, France, \\
										Email: n.gresset@fr.merce.mee.com}
										
   \IEEEauthorblockA{\IEEEauthorrefmark{4}%
                     Center for Pervasive Communications and Computing,
										 University of California, Irvine, CA,
                     USA,
                     Email:  syed@uci.edu} 
										
			\thanks{The  work  of  P.  de  Kerret  and  D.  Gesbert  was  supported  by  ERC  under Grant  670896. The collaboration with Syed A. Jafar and Arash G. Davoodi was carried out during the research visit of Antonio Bazco-Nogueras at the University of California at Irvine.}
}			
		\date{\today}
		\maketitle

	\begin{abstract}
		In this work, we study the impact of imperfect sharing of the Channel State Information (CSI) available at the transmitters	on a Network MIMO setting in which a set of $M$ transmit antennas, possibly not co-located, jointly serve two multi-antenna users endowed with $N_1$ and $N_2$ antennas, respectively. 
		We consider the case where only a subset of $k$ transmit antennas have access to perfect CSI, whereas the other $M-k$ transmit antennas have only access to finite precision CSI.  
	%
		The analysis of this configuration aims to answer the question of how much an extra informed antenna can help. 
		We model this scenario as a Decentralized MIMO Broadcast Channel (BC) and characterize the Degrees-of-Freedom (DoF) region, showing that only $k=\max(N_1,N_2)$ antennas with perfect CSI are needed to achieve the DoF of the conventional BC with ubiquitous perfect CSI. 
		Furthermore, we identify the increase of DoF obtained by providing CSI to an extra transmit antenna.
	\end{abstract}

		\section{Introduction}
		The availability of CSI at the Transmitters (CSIT) is one of the fundamental requirements for managing  interference  in Multiple-input Multiple-output (MIMO) and multi-user cooperative settings. 
		%
		On account of the infeasibility of acquiring perfect CSIT in many practical scenarios, there has been a significant interest in characterizing the impact of non-perfect CSIT on the system performance. 
		The non-perfect CSIT assumption has been analyzed from many different perspectives,  considering for example the cases of noisy instantaneous CSIT\cite{Davoodi2016_TIT_DoF}, perfect delayed CSIT\cite{MaddahAli2012}, partial \cite{Davoodi2019_ISIT}, hybrid \cite{Wang2017_TIT}, or alternating CSIT\cite{Tandon2012b,Rassouli2016}. 		
		However, it is normally assumed that the CSIT is centralized, i.e., \emph{perfectly shared} among the transmitters. 
		Although this belief arises naturally in MIMO settings with a single multi-antenna transmitter, it is unattainable in many practical settings with cooperative nodes or transmitters with remote radio-heads.  
		Such settings are expected to burgeon due to the increased heterogeneity and densification of the wireless networks. 

		Motivated by the foregoing, we aim to understand the impact of \emph{imperfectly shared} CSIT, i.e., the case in which each transmitter may have a different CSI. 
		This configuration, also known as \emph{Distributed CSIT} setting, has been previously studied in the literature for the Interference Channel with local CSI \cite{Aggarwal2012} or the Network MISO setting \cite{Bazco2018_WCL, Bazco2020_TIT}. 
		In this work, we focus on the Network MIMO setting. 
		Note that a Network MIMO setting in which the transmitters perfectly share the user data but not the CSIT can be modeled as a MIMO BC setting with antenna-dependent CSIT, and consequently we denote this setting as the Decentralized MIMO Broadcast Channel. 
				
		Therefore, we consider the 2-user MIMO BC where the users have $N_1$ and $N_2$ antennas, respectively. 
		The~DoF~metric of this setting has been analyzed for multiple heterogeneous,~yet centralized, CSI configurations; for example, the cases in which the CSIT for each user can be either Perfect, Delayed, or Not-available have been studied in\cite{Tandon2012_ISWCS,Wang2017_TIT, Tandon2012b}. 
		However, this work is to our knowledge the first to consider distributed CSIT. 
		In particular, we assume that only $k$ of the $M$ transmit antennas have access to perfect CSI, whereas the other $M-k$ transmit antennas have only access to finite precision CSI. 
		
		This model, in which some~transmit antennas are provided with global CSI (also from the other non-informed transmit antennas), arises in the context of FDD heterogeneous networks where the users feed back the global CSI to a main base station, which is in turn helped by secondary nodes or remote radio-heads with a limited backhaul.  
		The availability of the user data at all transmit antennas is feasible at the same time thanks to caching and Cloud/Fog-RAN technologies and the fact that, for many applications, the delay requirements for user data are several orders of magnitude slower than the channel coherence time\cite{Bazco2020_TIT}.   
			
		Our main contributions are as follows: $i)$ We present an outer bound for the DoF region of the 2-user MIMO BC when only $k$ transmit antennas have access to perfect CSI; 
		$ii)$ we show that having perfect CSIT at $k=\max(N_1,N_2)$ antennas is enough to achieve the DoF region  of the conventional  MIMO BC with perfect CSIT at every antenna; and $iii)$ we develop an achievable scheme that attains the DoF region for $k\geq \min(N_1,N_2)$ and partially closes the gap for  $k< \min(N_1,N_2)$. 

		\paragraph*{Notations}
		\noindent%
		For any matrix $\bA$, the element of the $i$-th row and $j$-th column of $\bA$ is denoted by $a_{ij}$. 
		Similarly, $\bA_{\left[x_1:x_2,y_1:y_2\right]}$ denotes the sub-matrix composed of the rows $\{x_1, x_1+1,\dots,x_2\}$ and the columns $\{y_1, y_1+1,\dots,y_2\}$ of $\bA$.
		$[n]$ is defined as $[n]\triangleq\{1,\dots,n\}$ and, in any variable $X$, the superscript $^{[n]}$ stands for $\left\{X(i)\right\}_{i\in[n]}$. 
		For any set of variables $\mathcal{S}$, $ H\big(\bigcap_{\substack{\Sc_i \in \Sc}} \Sc_i\big)$ denotes the joint entropy of the elements in $\mathcal{S}$.
	
	\section{System Model}
		
		\subsection{MIMO Broadcast Channel}
			
			We analyze the 2-user Network MIMO setting where $M$ transmit antennas (TXs) jointly serve 2~users (RXs) of $N_1$ and $N_2$ antennas, respectively. 
			We assume w.l.o.g. that $N_1 \leq N_2$. 
			The received signal at RX~$i$, $i\in\{1,2\}$, is given by 
			\eqm{
					\bY_i(t) \triangleq \sqrt{P}\,\bH_i (t)\bX (t) + \bN_i(t), \vspace{-2ex}
			}
			where $\bH_i \in \mathbb{R}^{N_i \times M}$ denotes the matrix of channel coefficients for RX~$i$, and the received signal vector is defined as $\bY_i(t) \triangleq \left[Y_{i,1}(t), Y_{i,2}(t),\dots,Y_{i, N_i}(t)\right]^\trans$. 
			The parameter~$t$ represents the channel use and~$P$ is the nominal SNR parameter.  
			We define the global channel matrix as
			\eqm{
					\bH^\trans \triangleq \left[\bH_1^\trans,\  \bH_2^\trans\right], \  \bH \in  \mathbb{R}^{(N_1 + N_2) \times M},
			}
			and the channel vector between TX~$j$ and RX$~i$ as $\bH_{i,j}$. 
			The transmit signal vector, which satisfies a unitary power constraint, is given by $\bX(t) \triangleq \left[X_1(t), X_2(t),\dots,X_M(t)\right]^\trans$. 
			$\bN_i(t)$ denotes the AWGN noise at RX~$i$. 
			RX~$i$ wants to receive a message $W_i$, and both $W_1$, $W_2$ are available at all the TXs.  
			The definitions of achievable rates $R_i(P)$ and capacity region $\Cc(P)$ are standard \cite{ElGamal2011}. 
			The DoF for RX$~i$ is defined as $d_i \triangleq \limpf \frac{R_i(P)}{\log \Pb}$, where $\Pb\triangleq\sqrt{P}$. 
			The closure of achievable DoF tuples $(d_1,d_2)$ is called the DoF region $\Dc$. 
					
		\subsection{Finite Precision CSI}
			We assume that the channel coefficients are bounded away from 0 and infinity and that  are drawn from distributions that satisfy the bounded density assumption, which is presented below. 	
		
			\begin{definition}[\!\!{\cite[Definition 4]{Davoodi2019_ISIT}}~Bounded Density Assumption]\label{def:bounded}
				Let $\Gc$ be a set of real-valued random variables which satisfies both of the following conditions.
				\enb
					\item The magnitudes of all the random variables in $\Gc$ are bounded away from infinity, i.e., there exists a constant $\Delta < \infty$ such that for all $g \in \Gc$ we have $|g|\leq \Delta$. 
					\item There exists a finite positive constant $f_{\max}$, such that for all finite cardinality disjoint subsets $\Gc_1, \Gc_2$ of $\Gc$, the joint probability density function of all random variables in $\Gc_1$, conditioned on all random variables in $\Gc_2$, exists and is bounded above by $f_{\max}^{|\Gc_1|}$.
				\ene
			\end{definition}~\vspace{-4.5ex}

			Consider a TX at which the channel coefficients are available only up to finite precision. For this TX, the channel coefficients satisfy the “bounded density assumption” of Definition~\ref{def:bounded}\cite{Davoodi2017_TIT_GDoF_IC_finite}.~\vspace{-1.5ex}

			
			\begin{figure}[t]\centering%
					
		\centering%
		\begin{tikzpicture}[thick,scale=0.3,>=stealth]
				\draw[] (0,0) node (X1){$X_1$};
				\draw[] (0,3) node (X2){$X_2$};
				\draw[] (0,8) node (X3){$X_3$};
				\draw[] (0,13) node (X4){$X_4$};

				\draw[] ([xshift=1.5cm, yshift=-1.5cm]X1.south) node (h){\large $~~~\bH$};
				\draw[->, line width =.5pt] ([xshift=1.5cm, yshift=-1.5cm]X1.south) -- ([xshift=-0cm, yshift=-1.5cm]X1.south) -- ([xshift=-0.0cm, yshift=-.7cm]X1.south); 
				
				\draw (-1.5,-1.5) rectangle (1.5,4.5);			
				\draw (-1.5, 6.5) rectangle (1.5,9.5);			
				\draw (-1.5,11.5) rectangle (1.5,14.5);		
								
				\draw[] (13,1) node (y1){$Y_{1,1}$};
				\draw[] (13,6) node (y2){$Y_{2,1}$};
				\draw[] (13,9) node (y3){$Y_{2,2}$};
				\draw[] (13,12) node (y4){$Y_{2,3}$};

				\draw[] (17,1) node (rx1){$\RX 1$};
				\draw[] (17,9) node (rx1){$\RX 2$};
				
				\draw (11.25,-0.5) rectangle (14.5,2.5);			
				\draw (11.25,4.5) rectangle (14.5,13.5);			
							
    \node[myAntennaTX, anchor=south west] at ([xshift=0.3cm]X1.east) {};
    \node[myAntennaTX, anchor=south west] at ([xshift=0.3cm]X2.east) {};
    \node[myAntennaTX, anchor=south west] at ([xshift=0.3cm]X3.east) {};
    \node[myAntennaTX, anchor=south west] at ([xshift=0.3cm]X4.east) {};
		
    \node[myAntennaRX, anchor=south east] at ([xshift=-0.3cm]y1.west) {};
    \node[myAntennaRX, anchor=south east] at ([xshift=-0.3cm]y2.west) {};
    \node[myAntennaRX, anchor=south east] at ([xshift=-0.3cm]y3.west) {};
    \node[myAntennaRX, anchor=south east] at ([xshift=-0.3cm]y4.west) {};
			
		\draw [dotted, line width=.5pt,  decorate,decoration={calligraphic brace,amplitude=8pt},xshift=-0pt,yshift=0pt] (-2.5,6.5) -- (-2.5,14.5) node [black,midway,xshift=-5.5ex] {%
				\begin{tabular}{c} ~\\[.75ex] ~~~$\bTX_\varnothing$  \\[.85ex] \footnotesize $(M-k = 2)$~~~ \end{tabular}%
			};
		\draw [dotted, line width=0.5pt, decorate,decoration={calligraphic brace,amplitude=7pt},xshift=-0pt,yshift=0pt] (-2.5,-1.5) -- (-2.5,4.5) node [black,midway,xshift=-5.5ex] {%
				\begin{tabular}{c}~\\[0.5ex] ~~~$\bTX_\star$  \\[.5ex] \footnotesize $~(k = 2)$ \end{tabular}%
			};
			
	\end{tikzpicture}~\vspace{-2ex}		
				\caption{System model for $(M,N_1,N_2,k) = (4,1,3,2)$. %
									The transmit antennas can belong to non-colocated transmitters. Note that  the antennas in $\bTX_\varnothing$ do not have access to $\bH$.
									} \label{fig:sys_mod}~\vspace{-6ex}
			\end{figure}
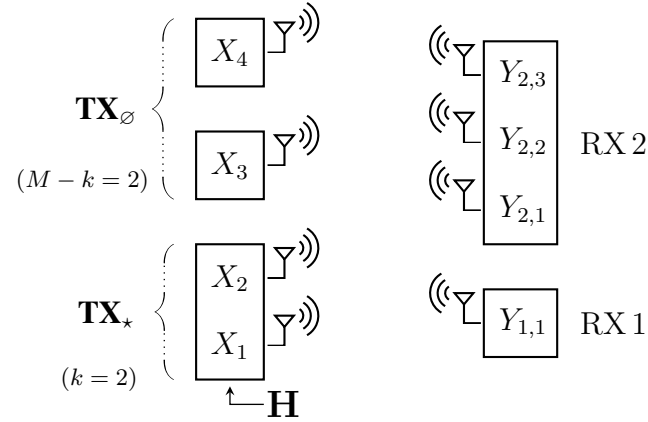

		\subsection{Distributed CSIT}
			We consider a Distributed CSIT setting\cite{Bazco2020_TIT} where the first $k$ TXs are provided with perfect CSI, such that they know the whole multi-user channel matrix~$\bH$, whilst the other $M-k$ TXs have only finite precision CSI.  
			We denote henceforth this setting as the $(M,N_1,N_2,k)$ MIMO~BC. 
			
			\begin{remark}
				The notation ``TX'' refers to a single transmit antenna. 		
				The transmit antennas can be distributed among an arbitrary number of physical transmitters. Thus, there can be for example $M$ single-antenna transmitters or two~$\frac{M}{2}$-antenna transmitters. 
			\end{remark}
			
			We split the set of transmit antennas in two different groups. 
			Let us first denote the $i$-th transmit antenna as $\TX_i$, $i\in[M]$.		
			Consequently, we define:
				\itb
					\item $\bTX_\star \triangleq \left[\TX_1,\dots, \TX_k\right]$ denotes the $k$ TXs that have access to perfect CSI, i.e., which know $\bH$ instantaneously. 
					\item $\bTX_\varnothing \triangleq \left[\TX_{k+1},\dots, \TX_M\right]$ denotes the $M-k$ TXs that have access only to finite precision CSI. 
					This implies that, for $\bTX_\varnothing$, the channel coefficients satisfy the bounded density assumption of Def.~\ref{def:bounded}~\cite{Davoodi2019_ISIT,Davoodi2017_ISIT_sumset}.
				\ite
			Similarly, we denote as $\bX_\varnothing$ (resp. $\bX_{\star}$) the transmit signal from $\bTX^{}_\varnothing$ (resp. $\bTX^{}_\star$).
					
			\begin{remark}	
				Although considering both perfect and finite CSI may resemble the conventional BC  with Hybrid CSIT in which there exists perfect CSIT for one RX and no CSIT for the other RX (the so-called `PN' setting)\cite{Wang2017_TIT,Hao2017,Davoodi2019_ISIT}, the CSI model here considered is substantially different: 
				In the mentioned `PN' setting, \emph{all} the TXs share the same CSI, i.e.,  all of them have access to perfect CSI for one RX and \emph{no} TX has access to  CSI of the other RX. 
				However, in our setting, a \emph{subset} of TXs has access to  perfect \emph{global} CSI (for both RXs), whereas the other subset has access only to finite precision CSI of the \emph{global} CSI. 
				Further discussion about this CSIT setting can be found in \cite{Bazco2020_TIT}.
			\end{remark}

		\section{DoF Region of the ($M,N_1,N_2,k$) Broadcast Channel}
		We analyze the DoF region of the MIMO BC as a function of the number of TXs with perfect CSIT ($k$). 
		Therefore, we can measure the gain (in terms of DoF) that is obtained by providing an extra TX with perfect CSIT, which would require either backhaul of feedback resources. We first present an outer bound. 

		%
			\begin{theorem}\label{theo:dof_r_big}
				Let us consider the $(M,N_1,N_2,k)$ MIMO BC. 
				If $k < N_2$ and $M > N_2$, the DoF region ($\Dc$) is enclosed in
					\begin{subequations}\label{eq:lem_reg_1}
						\begin{align}[left ={(d_1,d_2)	\in  \empheqlbrace}]
							& d_1 \leq \min(M,\; N_1)						\label{eq:lem_reg_1a} \\
							& d_2 \leq \min(M,\; N_2)						\label{eq:lem_reg_1b} \\
							&	d_1 + d_2 \leq \min(M,\; N_1+N_2) 	\label{eq:lem_reg_1c} \\
							& \frac{d_1}{\min(M,N_1+N_2)-k} + \frac{d_2 - k}{\min(N_2,M)-k} \leq 1	\label{eq:lem_reg_1d}						
					 \end{align}
					\end{subequations}~\vspace{.5ex}
														
				\noindent Otherwise (i.e., if $k \geq N_2$ or $M\leq N_2$), $\Dc$ is enclosed in 
				\begin{subequations}\label{eq:lem_reg_2}
						\begin{align}[left ={(d_1,d_2)	\in  \empheqlbrace}]
							& d_1 \leq \min(M,\; N_1)		\label{eq:lem_reg_2a} \\
							& d_2 \leq \min(M,\; N_2)		\label{eq:lem_reg_2b}\\
							&	d_1 + d_2 \leq \min(M,\; N_1+N_2) \label{eq:lem_reg_2c}
					 \end{align}
					\end{subequations}		
			\end{theorem} 
			
			\begin{IEEEproof}
				The proof is relegated to Section~\ref{se:converse_theo_dcsit}.
			\end{IEEEproof}

		The DoF region in~\eqref{eq:lem_reg_2} matches the DoF region of the MIMO BC with perfect CSIT.  
		Moreover, the bound~\eqref{eq:lem_reg_1d} holds for any value of $M$. However, note that, if $M \leq N_2$, \eqref{eq:lem_reg_1d} becomes~\eqref{eq:lem_reg_2c}, and hence we recover~\eqref{eq:lem_reg_2}. 
		Let us consider now the sum DoF, defined as $d_\Sigma \triangleq \max_{(d_1,d_2)\in\Dc} (d_1+d_2)$. 

			\begin{lemma}\label{lem:s_dof_big}
				The sum DoF of the $(M,N_1,N_2,k)$ MIMO BC is upper-bounded by
					\eqm{
						d_\Sigma & \leq \min\Big( N_1+N_2,\ M,\ 
							N_2 +\frac{ N_1\min(N_1,M-N_2)}{\min(N_1+N_2,M)-k}\Big).
					}
			\end{lemma}
		Lemma~\ref{lem:s_dof_big} follows as a direct aftermath of Theorem~\ref{theo:dof_r_big}. 
		Thus, the sum DoF upper bound is strictly smaller than the DoF of the BC with perfect CSIT\cite{Wang2017_TIT} for the regime of~\eqref{eq:lem_reg_1}, and  matches it for the regime of~\eqref{eq:lem_reg_2}. 
		Next, we introduce the achievability results.  
			\begin{theorem}\label{theo:achiev_dof_region}
				The DoF region outer bound of Theorem~\ref{theo:dof_r_big} is achievable for $k\geq \min(N_1,N_2)$. 
			\end{theorem}
				\begin{IEEEproof}
					The proof follows from a novel transmission scheme introduced in Section~\ref{se:dof_achiev}, which shows that the sum DoF of Lemma~\ref{lem:s_dof_big} is achievable. 
					The DoF region can be obtained then by time-sharing. 
					The transmission scheme achieving Theorem~\ref{theo:achiev_dof_region} is based on the Active-Passive Zero-Forcing precoding (AP-ZF) introduced in~\cite{Bazco2020_TIT} and the fact that exploiting the unavoidable interference as side information is beneficial.   
				\end{IEEEproof}			
			\noindent Hence, for the simple case where $M = N_1 + N_2$, it follows that 
				\begin{align}
					d_\Sigma = \begin{cases} 
													N_2 + N_1 & \text{if $k \geq N_2$}\\
													N_2 + N_1\frac{N_1}{N_1+N_2-k} & \text{if $N_1 \leq k < N_2$}.
										\end{cases}
				\end{align}
		Unfortunately, besides particular cases, no tight general bound is known for the regime  $k < N_1$. 
		Nevertheless, we can extend the proposed scheme to obtain a general lower bound, which is stated in the following proposition. 
		
			\begin{proposition}\label{prop:achiev_m_min_bad}
				Let us assume that $k<N_1$. 
				Then, the sum DoF of the  $(M,N_1,N_2,k)$ MIMO BC is lower-bounded by			
				\eqmo{
						d_\Sigma
							& \geq 
									\max\Big(\min(N_2,M),\ \min(N_2,M-k) + \frac{k^2}{\min(N_2,M-k)}\Big).
				}			
			\end{proposition}			
				\begin{IEEEproof}
					The proof is relegated to Section~\ref{se:case_m_min}. 
				\end{IEEEproof}
				
		\section{Discussion}
		The sum DoF of the 2-user MIMO BC with perfect CSIT is $\DoF^\star = \min(M,N_2 + N_1)$\cite{Wang2017_TIT}. 
		Hence, Theorem~\ref{theo:dof_r_big} implies that we only need perfect CSI at $k = N_2$ to recover the maximum DoF. 
		This aftermath extends the results of previous works on the MISO setting\cite{Bazco2018_WCL,Bazco2020_TIT}, where it was shown that having the most accurate CSI at only a subset of TXs is (sometimes) enough to recover the DoF achieved with perfect CSI sharing. 
		%
		
		Fig.~\ref{fig:dof_m} represents the sum DoF as a function of $k$. 
		We observe how for $k\geq N_2$ the DoF obtained with centralized perfect CSIT is attained, and that for $N_1\leq k \leq N_2$ the bound is tight. 
		For the case $k<N_1$, there exists a gap between the upper and the lower bound. 
		We can infer that the upper bound is loose from the fact that for $k=0$ we obtain that $\DoF = N_2 + 1$, whereas it is known that the DoF of the MIMO BC with finite precision CSIT is $\DoF = N_2$\cite{Davoodi2019_ISIT}.  
		It is noteworthy that, the closer $k$ is to the number of antennas of any of the RXs, the more the DoF increases from $k$ to $k+1$. 
		In Fig.~\ref{fig:dof_region}, we present the DoF region for the case~$(4,1,3,k)$. 
		Interestingly, a single informed antenna can considerably increase the performance, specially for RX~1. 
		
			\begin{figure}[t]\centering%
					
	\makeatletter
	\def\markxof#1{
		\pgf@process{#1}
		\pgfmathparse{
			\pgf@x/\pgfplotsunitxlength+
			\pgfplots@data@scale@trafo@SHIFT@x)/10^\pgfplots@data@scale@trafo@EXPONENT@x}
	}
	\makeatother

	\pgfplotsset{
		mystyle/.style={%
			axis x line=bottom,
			axis y line=left,
			axis y discontinuity = parallel,
			xmin=0, xmax=9.75,
			samples =13,
			ymax = 9.75, 
			ymin  = 3,
			xticklabels={},
			extra x ticks={0,3,6,9},
			extra x tick labels={0,$N_1 =3$,$N_2 = 6$,$M =9$},			
			yticklabels={},
			extra y ticks={6,7.5,9},
			unit vector ratio*=1 1 1,
			legend style={
					draw=none,
					at={(0.05,0.2)},
					anchor=west,
			},				
		}
	}

	\begin{tikzpicture}
		\begin{axis}[
			mystyle,
			xlabel={$k$},
			ylabel={$\DoF$},
			xlabel style={at={(1,0.125)}, anchor=south},
			ylabel style={rotate=-90, at={(0.01,1)}, anchor=west},
			legend cell align=left,
		]

			\addplot[name path=L, opacity=0, mark=none, domain=0:3, forget plot ] {F_upper};
			\addplot[name path=line, opacity=0,  no markers, domain=0:3,  forget plot ] {F_lower};
			\addplot[gray!30, forget plot ] fill between[ 
				of = L and line, soft clip={domain=0:3}
			];

			\addplot[myblue, smooth, domain=0:6, line width=0.75pt] {6+(9)/(9-\x)} ;
			 \addlegendentry{Upper bound}

			\addplot[mygreen, smooth, domain=0:3, only marks, samples = 4, mark size=2pt, line width=0.75pt, line width=1pt]{6+(\x*\x)/6} ;
			 \addlegendentry{Lower bound}

			\addplot[gray!30] fill between[ 
				of = L and line, soft clip={domain=0:3}
			];
			\addlegendentry{Gap}
			
			\addplot[myblue, smooth, domain=6:9, line width=0.75pt] {9} ;

			\addplot[mygreen, smooth, domain=3:6, only marks, samples = 4, mark size=2pt, line width=0.75pt, line width=1pt]{6+9/(9-\x)} ;
			\addplot[mygreen, smooth, domain=6:9, only marks, samples = 4, mark size=2pt, line width=0.75pt, line width=1pt]{9} ;
			
			\addplot[black, thick, dashdotted] coordinates {(0,6) (9,6)};
			\addplot[black, thick, dashed] coordinates {(0,9) (9,9)};	
					
			\addplot[black, densely dotted] coordinates {(0,7.5) (3,7.5) (3,0)};
			\addplot[black, densely dotted] coordinates {(0,9) (6,9) (6,0)};
			\addplot[black, densely dotted] coordinates {(9,9) (9,0)};
			
			\node[label={0:{DoF for no CSIT}}, inner sep=0pt, anchor=south] at (axis cs:4.25,6.3) {};
			\node[label={0:{DoF for Perfect CSIT}}, inner sep=0pt] at (axis cs:2,9.4) {};							
							
		\end{axis}

	\end{tikzpicture} ~\vspace{-1.5ex}
				\caption{Sum DoF as a function of the number of transmit antennas with perfect CSIT ($k$) for the case ($M,N_1,N_2$) = ($9,6,3$).} \label{fig:dof_m}~\vspace{-2.5ex}	
			\end{figure}
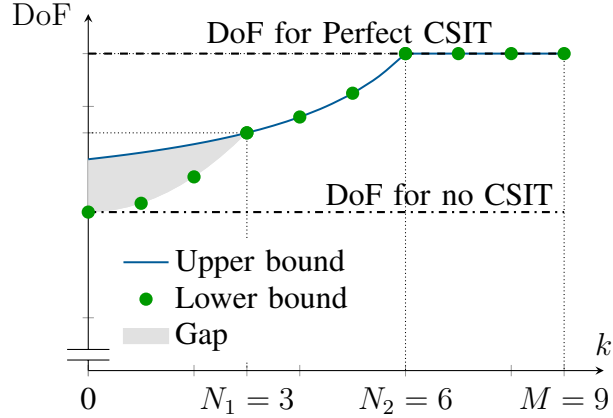
		
			\begin{figure}[t]\centering%
					
	\makeatletter
	\def\markxof#1{
		\pgf@process{#1}
		\pgfmathparse{
			\pgf@x/\pgfplotsunitxlength+
			\pgfplots@data@scale@trafo@SHIFT@x)/10^\pgfplots@data@scale@trafo@EXPONENT@x}
	}
	\makeatother

	\pgfplotsset{
		mystyle/.style={%
			clip=true,
			axis x line=bottom,
			axis y line=left,
			xmin=0, xmax=3.75,
			samples =32,
			ymax = 1.75, 
			ymin  = 0,
			xticklabels={},
			extra x ticks={3},
			extra x tick style={xticklabel=\pgfmathprintnumber{\tick}},	
			yticklabels={},
			extra y ticks={1},
			extra y tick style={yticklabel=\pgfmathprintnumber{\tick}},
			unit vector ratio*=1 1 1,
			legend columns=2, 
			legend style={
					draw=none,
					at={(0.01,0.8)},
					anchor=west,
					/tikz/column 2/.style={
							column sep=5pt,
					},
			},				
		}
	}

	\begin{tikzpicture}
		\begin{axis}[
			mystyle,
			xlabel={$d_2$},
			ylabel={$d_1$},
			xlabel style={at={(1,0.14)}, anchor=south},
			ylabel style={rotate=-90, at={(0.075,1)}, anchor=west},
			legend cell align=left,
		]

			\addplot[mygreen, line width=2.5pt, label={$k=3$}]  coordinates {(0,1) (3,1) (3,0)};
			 \addlegendentry{$k=3$}
			\addplot[myblue, line width=1.5pt] coordinates {(3,0) (5/2,1)};
			 \addlegendentry{$k=2$}			
			\addplot[red, dashed, line width=1.5pt] coordinates {(3,0) (7/3,1)};
			 \addlegendentry{$k=1$}			
			\addplot[black, thick, densely dotted] coordinates {(3,0) (0,1)};
			 \addlegendentry{$k=0$}

			\node[label={0:{\color{mygreen}(1,3)}},circle, draw=black, fill=mygreen, line width=.75pt, inner sep=2.05pt] at (axis cs:3,1) {};
			\node[label={80:{\color{myblue}$(1,\frac{5}{2})$}},circle, draw=black, fill=myblue, line width=.75pt, inner sep=2.05pt] at (axis cs:2.5,1) {};
			\node[label={200:{\color{red}$(1,\frac{7}{3})$}},circle, draw=black, fill=red, line width=.75pt, inner sep=2.05pt] at (axis cs:7/3,1) {};
		\end{axis}

	\end{tikzpicture} ~\vspace{-1.5ex}
				\caption{DoF region for the ($M,N_1,N_2,k$) = ($4,1,3,k$) MIMO BC with $k\in\{0,1,2,3\}$.} \label{fig:dof_region}~\vspace{-1.5ex}
			\end{figure}
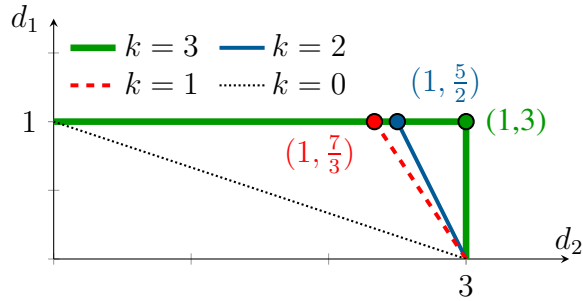

			\begin{figure}[t]\centering%
					
	\makeatletter
	\def\markxof#1{
		\pgf@process{#1}
		\pgfmathparse{
			\pgf@x/\pgfplotsunitxlength+
			\pgfplots@data@scale@trafo@SHIFT@x)/10^\pgfplots@data@scale@trafo@EXPONENT@x}
	}
	\makeatother

	\pgfplotsset{
		mystyle/.style={%
			axis x line=bottom,
			axis y line=left,
			axis y discontinuity = parallel,
			xmin=10, xmax=20.75,
			samples =13,
			ymax = 21.5, 
			ymin  = 9,
			xticklabels={},
			extra x ticks={10,12,16,19},
			yticklabels={},
			extra y ticks={10,18,20},
			legend style={
					draw=none,
					at={(0.825,0.15)},
					anchor=east,
			},				
		}
	}

	\begin{tikzpicture}
		\begin{axis}[
			mystyle,
			width=0.55\textwidth,
			xlabel={$N_2$},
			ylabel={$\DoF$},
			xlabel style={at={(1.0,0.095)}, anchor=west},
			ylabel style={rotate=-90, at={(0.075,.925)}, anchor=south},
			legend cell align=right,
		]

			\addplot[myblue, smooth, domain=10:12, line width=0.75pt, forget plot] {20-0*\x} ;
			\addplot[myblue, smooth, domain=12:19, line width=0.75pt] {\x+(20-\x)*(20-\x)/8} ;
			 \addlegendentry{Upper bound}

			\addplot[mygreen, smooth, domain=10:12, only marks, samples = 3, mark size=2pt, line width=0.75pt, line width=1pt, forget plot]{20} ;			
			\addplot[mygreen, smooth, domain=12:19, only marks, samples = 8, mark size=2pt, line width=0.75pt, line width=1pt]{\x+(20-\x)*(20-\x)/8} ;
			 \addlegendentry{Lower bound}
			
			\addplot[black, thick, domain=10:19, dashdotted] {\x} ;
			\addplot[black, thick, domain=10:19, dashed] {20} ;

			\node[label={[rotate=35.5]0:{DoF for no CSIT}}, inner sep=0pt, anchor=south, rotate=30] at (axis cs:12,11.75) {};
			\node[label={[anchor=center]0:{DoF for Perfect CSIT}}, inner sep=0pt] at (axis cs:16,20.8) {};				
			\node[label={[anchor=west]0:{\small $19.125$}}, inner sep=0pt] at (axis cs:19,19.4) {};				
					
			\addplot[black, densely dotted] coordinates {(10,18) (16,18) (16,5)};
			\addplot[black, densely dotted] coordinates {(12,18) (12,5)};
			\addplot[black, densely dotted] coordinates {(19,20) (19,5)};
							
		\end{axis}

	\end{tikzpicture}~\vspace{-8ex} 
				\caption{DoF as function of $N_2$ for the setting ($M,k$) = ($20,12$) and $N_2+N_1=M$.  }\label{fig:dof_n2} %
			\end{figure}
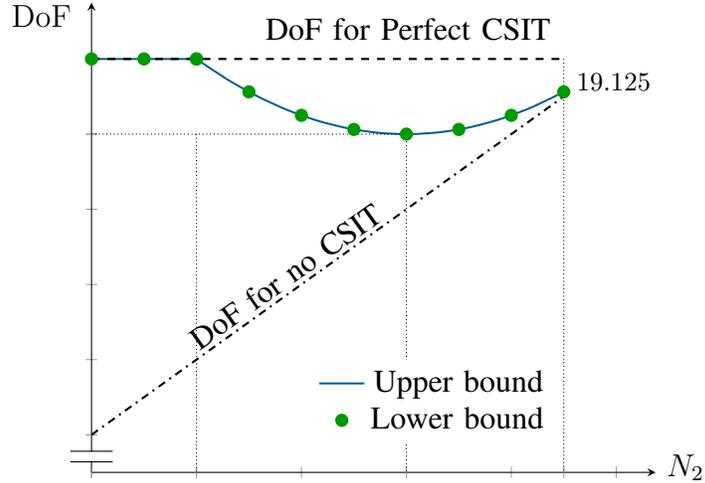
				
		Finally, Fig.~\ref{fig:dof_n2} illustrates the DoF as a function of the \emph{repartition} of antennas among the RXs, i.e., for a fixed-size setting with $N_1+N_2 = M = 20$ and $k=12$ transmit antennas with perfect CSI, we plot the DoF as a function of $N_2$.  
		Therefore, $N_1$ is obtained as $20-N_2$.

		Besides this, the DoF expression obtained for this decentralized setting has an appreciable similarity with the DoF expression of the centralized MIMO BC in which the transmitter has perfect CSI for RX~1 and delayed CSI for RX~2, also known as the `PD' setting\cite{Tandon2012_ISWCS}. 	
		The DoF region of the  `PD' setting was derived in~\cite{Tandon2012_ISWCS}. 
		Let us recall that the sum DoF of the $(M,N_1,N_2,k)$ MIMO BC is denoted by $d_\Sigma$. Similarly, the sum DoF  of the centralized `PD' setting is denoted by $d_\Sigma^{\PD}$, such that  
			\eqm{
					d_\Sigma^{\PD} & \triangleq d_1^{\PD} + d_2^{\PD}, 
			}%
			\FloatBarrier%
		\noindent where $d_i^{\PD}$ stands for the DoF of RX~$i$ in the centralized `PD' setting. 
		By way of example, consider the scenario in which the number of transmit antennas~($M$) is the same as the sum of receive antennas, i.e.,  $M = N_1 + N_2$. 
		Furthermore, suppose that the number of transmit antennas with perfect global CSI ($k$) satisfies that $N_1 \leq k < N_2$.
		This assumption is made so as to consider the particular bound of~\eqref{eq:lem_reg_1d}, 
			\eqm{
					\frac{d_1}{\min(M,N_1+N_2)-k} + \frac{d_2 - k}{N_2-k} \leq 1.
			}
		The upper bound for the `PD' case was derived by R.~Tandon \emph{et al.} in~\cite{Tandon2012_ISWCS} and, for the case with $M=N_1+N_2$, it writes as
			\eqm{
				\frac{d_1^{\PD}}{N_1+N_2} + \frac{d_2^{\PD}}{N_2} \leq 1.
			}
		This weighted expression leads to a sum DoF of 
			\eqm{
				d_\Sigma^{\PD}  &= N_1 \ \  + \ \  N_2  \ - \  N_2\frac{N_1}{N_1+N_2}.
			}
		On the other hand, from Lemma~\ref{lem:s_dof_big} it follows that $d_\Sigma$ is given by 
			\eqm{
				~~~~~d_\Sigma  &= N_1\ \   +\ \  N_2  \ - \  \frac{(N_2-k)N_1}{N_1+(N_2-k)}. \label{eq:dof_distr_achiev}
			}
		If we compare these two settings with the perfect-CSIT MISO BC, we can observe that there exists an analogy between both settings:
			\enb \setlength\itemsep{0em}
					\item In the `PD' setting, the loss of DoF due to having delayed CSIT for RX~2 instead of perfect CSIT is $- N_2\frac{N_1}{N_1+N_2}$. 
					\item In our decentralized setting, the loss of DoF due to having perfect CSIT \emph{only} at $k$ antennas is $- (N_2-k)\frac{N_1}{N_1+(N_2-k)}$. 
			\ene
		Therefore, the  $(M,N_1,N_2,k)$ setting seems analogous to a `PD' case where only $N_2-k$ antennas suffer from having delayed CSI instead of perfect CSI. 
				%
		An intuition behind this result is that, in our setting, we can apply a change of basis at RX~2 so that the TXs with perfect CSI ($\bTX_\star$) are only listened by $k$ antennas of RX~2. 
		Hence, even if those TXs have perfect CSI for the other $N_2-k$ antennas, those  antennas receive only information from the TXs with finite precision CSI ($\bTX_\varnothing$). 
				
								
		\section{Converse of Theorem~\ref{theo:dof_r_big}}\label{se:converse_theo_dcsit}
		We prove Theorem~\ref{theo:dof_r_big} for real channels. 
		The extension to complex variables is intuitive but cumbersome, and hence we omit it for sake of conciseness. 
		First, let us consider a genie-aided setting with perfect CSIT available at every transmit antenna. 
		This genie-aided scenario corresponds to the well-known conventional MIMO BC with perfect CSIT\cite{Wang2017_TIT}, whose DoF region coincides with~\eqref{eq:lem_reg_2}. 
		Since providing with additional CSI can not hurt, we obtain that~\eqref{eq:lem_reg_2} is an outer bound for the $(M,N_1,N_2,k)$ MISO BC.  
		Hence, it remains to prove that the bound \eqref{eq:lem_reg_1d}, 
			\eqm{
				\frac{d_1}{\min(M,N_1+N_2)-k} + \frac{d_2 - k}{N_2-k} \leq 1,\label{eq:bound_dcsit_e}
			}
		holds when $M > N_2$ and $k < N_2$. 
		Hence, we consider only the case in which $M > N_2$ and $k < N_2$. 
		We split the proof in two sub-regimes: $N_2 < M\leq N_1+N_2$ and $M> N_1+N_2$. 
		The regime in which $N_2 < M\leq N_1+N_2$ is considered below, whereas the outer bound for the case $M> N_1+N_2$ follows from invertible transformations at the nodes and is relegated to the Appendix. 
		
			\subsection{Converse for the case $N_2<M\leq N_1+N_2$}\label{se:region_outerbound_a}
				\subsubsection{Deterministic Channel Model}\label{subsubse:determinst_chan}
				We start similarly as in \cite{Davoodi2016_TIT_DoF,Davoodi2019_ISIT,Davoodi2017_TIT_GDoF_IC_finite} by discretizing the channel, what leads to a deterministic channel model introduced in\cite{Bresler2008}. 
				The discretized model is such that the input signals $\Xbar_j(t)\in\Zb$ and output signals $\Ybar_i(t)\in\Zb$ are given by
					\eqm{
						\Xbar_j(t) &\in \{0,1,\dots,\lceil\Pb\rceil\}, \quad \forall j \in [M], \\
						\bYbar_i(t) &\triangleq \sum_{j=1}^{M} \lfloor \bH_{i,j}\Xbar_j(t)\rfloor,		\quad		\forall i \in\{1,2\}.
					}
				In the following, we obtain an outer bound for this channel model. 
				From~\cite[Lemma~1]{Davoodi2016_TIT_DoF}, this DoF outer bound is also an outer bound for the channel model that we have considered.
				
			\subsubsection{Weighted sum rate}\label{subsubse:determinst_zeechan}		
			We obtain~\eqref{eq:bound_dcsit_e} by means of bounding the weighted sum rate $n(N_2-k)R_1 + n(M-k)R_2$. 
			First of all, we present an instrumental lemma. 
			\begin{lemma}\label{lem:lemma_bound}
					Let the number of transmit antennas with perfect CSIT satisfy  that $k <  N_2$. Then, 
					\eqmo{
							&(N_2-k) H(\bYbar^{[n]}_{1}\mid \bH\nt,W_2) - (M-k) H(\bYbar^{[n]}_{2}\mid\bH\nt,W_2) \leq o(\log \Pb). \label{eq:lemma_reg_bound} 
					}
			\end{lemma}
				\begin{IEEEproof}
					The proof is relegated to Section~\ref{se:proof_lemma_ch5_bound}.
				\end{IEEEproof}
			We start  from Fano's inequality to obtain 
				\eqmo{
						 n(N_2-k)R_1 + n(M-k)R_2&\leq (M-k)  I(W_2;\bYbar^{[n]}_2\mid \bH\nt ) \\
							& \quad   \ + (N_2-k) I(W_1;\bYbar^{[n]}_1| \bH\nt, W_2) \\
								&\leq (M-k) \big( H(\bYbar^{[n]}_{2}\mid\bH\nt) - H(\bYbar^{[n]}_{2}\mid\bH\nt,W_2)\big) \\
							& \quad   \ + (N_2-k) H(\bYbar^{[n]}_{1}\mid\bH\nt,W_2) + o(n).  
				}
			The entropy of a random variable is bounded by its support, i.e., 
					$H(\bYbar^{[n]}_{2})\leq  N_2\,n\log\Pb$. 
			This fact and Lemma~\ref{lem:lemma_bound}   yield
				\eqmo{
					n(N_2-k)R_1 + n(M-k)R_2
						& \leq n \, (M-k) N_2  \log \Pb + n\,o(\log \Pb) + o(n). \label{eq:bound_reg_proof_1}
				}
			We can divide by~$(M-k)(N_2-k)$ to write
				\eqm{
					\frac{n R_1}{M-k} + \frac{n R_2}{N_2-k} &\leq \frac{n N_2  \log \Pb}{N_2-k} + n\, o(\log \Pb) + o(n). \label{eq:bound_reg_endsdff_3}							
				}				
			From the definition of DoF,  
			it follows that
				\eqmo{
					\frac{d_1}{M-k} + \frac{d_2}{N_2-k} &\leq \frac{N_2}{N_2-k}  
					\quad\Rightarrow\quad \frac{d_1}{M-k} + \frac{d_2-k}{N_2-k} &\leq 1,  \label{eq:bound_reg_end_4}
				}
			what concludes the proof of~\eqref{eq:lem_reg_1d} for $N_2< M \leq N_1+N_2$. \qed

			\subsection{Proof of Lemma \ref{lem:lemma_bound}}\label{se:proof_lemma_ch5_bound}

			We split the proof of Lemma \ref{lem:lemma_bound} in several steps. 
			First, we present some required definitions and lemmas, and we introduce some notation to explicitly show the dependence of the received signals on the input signals. 
			Second, we prove the key step for the proof, which is based on the sub-modularity property of the entropy. 
			To conclude, we explain how we can obtain Lemma \ref{lem:lemma_bound} by handily repeating the previous key step. 
			
			\subsubsection{Preliminary steps}\label{se:proof_lemma_ch5_bound_a}
		
			Let us recall a key definition from~\cite{Davoodi2017_ISIT_sumset}.
				\begin{definition}[\!\!{\cite[Def. 4]{Davoodi2017_ISIT_sumset}}]\label{def_L_Lb}
					For real numbers $x_1,x_2,\dots,x_K$, define the notations $L^b_j(x_i,\ i\in[K])$, and $L_j(x_i, i\in[K])$, as 
						\eqm{ 
								L^b_j(x_1,x_2,\dots,x_k) &\triangleq \sum\nolimits_{i\in[K]}\lfloor g_{j,i}x_i\rfloor \\
								L_j(x_1,x_2,\dots,x_k) &\triangleq \sum\nolimits_{i\in[K]}\lfloor h_{j,i}x_i\rfloor
						}
					for distinct random variables $g_{j,i}\in\Gc$ satifying the bounded density assumption, and for some arbitrary real valued and finite constants $h_{j,i}\in\Hc$, $|h_{j,i}|\leq \delta_z < \infty$. 
					The subscript $j$ is used to distinguish among multiple sums. 
				\end{definition}
			We recall that $\bYbar^{[n]}_{i} \triangleq [\Ybar^{[n]}_{i,1},\dots, \Ybar^{[n]}_{i,N_i}]$. 
			Moreover, it follows from Definition~\ref{def_L_Lb} that we can write $\Ybar_{i,j}(t)$ as $\Ybar_{i,j}(t) = L_{i,j}(t)(\Xbar_1(t),\dots,\Xbar_M(t))$. 
			Note that the signals $\Xbar\nt_1,\dots, \Xbar\nt_k$ may be a function of the messages \emph{and the channel}, but $ \{\Xbar_{k+1}\nt,\dots, \Xbar_M\nt\}\triangleq\bXbar_{\varnothing}\nt$ are independent of the channel.  
			We can apply a rotation matrix at RX~2 such that the $k$ first TXs ($\bTX^{}_\star$) are only heard by the first $k$ antennas of RX~2. 
			Hence, for any $k<j\leq N_2$, we have that
				\eqm{
						\Ybar^{[n]}_{2,j} = L^{b[n]}_{\Ybar,j}(\bXbar\nt_{\varnothing}).  \label{eq:dep_m_tre_2}
				}
			Thus, for any $k<j\leq N_2$, the coefficients of the linear combination $L^{b[n]}_{\Ybar,j}$ satisfy the bounded density assumption of Definition~\ref{def:bounded} because $\bTX_\varnothing$ has only finite precision CSI. 
			We omit hereinafter that $j\leq N_2$ for ease of readability. 
			From the fact that $H(A,B)\geq H(A)$, we obtain the following inequality. 
				\eqmo{
					& (N_2-k) H(\bYbar^{[n]}_{1}\mid\bH\nt,W_2) - (M-k) H(\bYbar^{[n]}_{2}\mid\bH\nt,W_2)\\
						& \hspace{20ex} \leq 	(N_2-k) H(\bYbar^{[n]}_{1}\mid\bH\nt,W_2) - (N_2-k)H\big(\bYbar^{[n]}_{2} \mid\bH\nt,W_2\big)\\ 
						& \hspace{20ex}\quad\ - (M-N_2) H\big(\bigcap_{j>k} \ \Ybar^{[n]}_{2,j} \mid\bH\nt,W_2\big). \label{eq:lemma_proof_zef} 
				}	
			From~\eqref{eq:dep_m_tre_2}, we can write that
				\eqm{
						H\big(\bigcap_{j>k}\Ybar^{[n]}_{2,j}\mid\bH\nt,W_2\big) = H\big(\bigcap_{\mathclap{j>k}} L^{b[n]}_{\Ybar,j}(\bXbar\nt_{\varnothing}) \mid\bH\nt,W_2\big),\label{eq:sdsfo}
				}
			which shows that, for any $j>k$, $\Ybar^{[n]}_{2,j}$ only depends on the $M-k$ input signals that form $\bXbar\nt_{\varnothing}$. 
			
			Let us first describe the intuition behind the proof before deriving the result.
			In~\eqref{eq:lemma_proof_zef}, there are  $N_2-k$~negative entropy terms, each one of $N_2$ variables, and another $M-N_2$~negative entropy terms, each one of $N_2-k$ variables. 
			All the variables are linear combinations of the $M$ transmit signals ($\Xbar_i$). 
			Our goal is to show that all those negative terms can be reordered so as to create $N_2-k$~terms of $M$ independent linear combinations. 
			If this statement is true, from the fact that $H(A) - H(B)\leq H(A|B)$, we can remove the contribution of the  $N_2-k$ positive terms $H(\bYbar^{[n]}_{1}|\bH\nt,W_2)$, since we can decode the $M$ signals with high probability from $M$ independent linear combinations. 
			In the following we show rigorously that the previous idea is indeed applicable. 
			For that purpose, we next present the fundamental step that allows us to \emph{reorder} the entropy terms. Later, we show how this step can be properly repeated so as to prove  Lemma \ref{lem:lemma_bound}.
		
			We further present a useful lemma that follows directly from \cite{Davoodi2016_TIT_DoF}. 
				\begin{lemma}\label{lem:lemma_finite_csit_z}
					Consider $\beta>0$ and random variables $F^{[n]}_j, G^{[n]}_j,\ j \in [J]$ that satisfy the bounded density assumption. 
					Let $\Xbar^{[n]}_j$ be independent of  $F^{[n]}_j, G^{[n]}_j$, for any $j \in [J]$.
					Then, it holds that
					\eqm{
							H\big(\sum_{j=1}^{J}\lceil \Pb^\beta F^{[n]}_j \Xbar^{[n]}_j\rceil\big) \leq H\big(\sum_{j=1}^{J}\lceil \Pb^\beta G^{[n]}_j \Xbar^{[n]}_j\rceil\big) + o(\log \Pb). \nonumber%
					}%
				\end{lemma}%
			\subsubsection{Applying the sub-modularity property}\label{se:proof_lemma_ch5_bound_b}
			
			First, let us note that we can re-write the received signal vector $\bYbar^{[n]}_{2}$ by applying~\eqref{eq:dep_m_tre_2} so as to obtain that
				\eqmo{\label{eq:y2_trans}
						\bYbar^{[n]}_{2} \triangleq \{\bigcap_{m\leq k}\Ybar^{[n]}_{2,m},\ \bigcap_{j>k} L^{b[n]}_{\Ybar,j}(\bXbar\nt_{\varnothing})\}.			
				} 	
			Hereinafter, we omit the $o(\log \Pb)$ terms for ease of notation and because they are irrelevant for the DoF metric. 					
			Lemma~\ref{lem:lemma_finite_csit_z} and the fact that $H(L(X_i))\leq H(L^b(X_i))$\cite{Davoodi2017_TIT_GDoF_IC_finite,Davoodi2017_ISIT_sumset} yield 
				\eqmo{
					&H\big(\bigcap_{j>k} L^{b[n]}_{\Ybar,j}(\bXbar\nt_{\varnothing})\mid\bH\nt,W_2\big) \geq H\big(L^{[n]}(\bXbar\nt_{\varnothing}),\ \bigcap_{\mathclap{j>k+1}} L^{b[n]}_{\Ybar,j}(\bXbar\nt_{\varnothing})  \mid\bH\nt, W_2\big),\label{eq:proof_converse_3e}
				}
			%
			In order to bound~\eqref{eq:lemma_proof_zef}, we first consider the term  $H\big(\bYbar^{[n]}_{2} |\bH\nt,W_2\big) + H\big(\bigcap_{j>k} \Ybar^{[n]}_{2,j} |  \bH\nt,W_2\big)$, which~appears in the negative terms of~\eqref{eq:lemma_proof_zef}. 
			It follows that
				\eqmo{
					& H\big(\bYbar^{[n]}_{2} \mid\bH\nt,W_2\big) + H\big(\bigcap_{j>k} \Ybar^{[n]}_{2,j} \mid\bH\nt,W_2\big) \\
						& \hspace{3ex}\  \ \myoverset{(a)}{=} H\big(\!\!\bigcap_{m\leq k}\!\Ybar^{[n]}_{2,m}, \bigcap_{j>k}L^{b[n]}_{\Ybar,j}(\bXbar_{\varnothing}) \mid\bH\nt,W_2\big) +  H\big(\bigcap_{j>k} L^{b[n]}_{\Ybar,j}(\bXbar_{\varnothing})\mid\bH\nt,W_2\big)\\
						& \hspace{3ex} \  \ \myoverset{(b)}{\geq}  H\big(\!\!\bigcap_{m\leq k}\!\Ybar^{[n]}_{2,m}, \!\bigcap_{j>k}\! L^{b[n]}_{\Ybar,j}(\bXbar\nt_{\varnothing}) \mid\bH\nt,W_2\big) + H\big(\bigcap_{\mathclap{j>k+1}} L^{b[n]}_{\Ybar,j}(\bXbar\nt_{\varnothing}), L^{[n]}(\bXbar\nt_{\varnothing})\mid\bH\nt,W_2\big)\\
						& \hspace{3ex} \  \  \myoverset{(c)}{\geq} H\big(\!\!\bigcap_{m\leq k}\!\Ybar^{[n]}_{2,m},\! \bigcap_{j>k}\! L^{b[n]}_{\Ybar,j}(\bXbar_{\varnothing}), L^{[n]}(\bXbar_{\varnothing}) \mid\bH\nt,W_2\big)  + H\big(\bigcap_{\mathclap{j>k+1}} L^{b[n]}_{\Ybar,j}(\bXbar\nt_{\varnothing})\mid\bH\nt,W_2\big) 
						\\
						& \hspace{3ex} \  \  \myoverset{(d)}{\geq} H\big(\bYbar^{[n]}_{2} , L^{[n]}(\bXbar\nt_{\varnothing}) \mid\bH\nt,W_2\big)  +  H\big(\bigcap_{\mathclap{j>k+1}} L^{b[n]}_{\Ybar,j}(\bXbar\nt_{\varnothing})\mid\bH\nt,W_2\big), \label{eq:bound_dd}
				}
			where $(a)$ follows from~\eqref{eq:dep_m_tre_2} and \eqref{eq:y2_trans}, $(b)$ from~\eqref{eq:proof_converse_3e}, $(c)$ comes from the sub-modularity property, which states that~$H(A,B)+H(B,C)\geq H(A,B,C) + H(B)$\cite[Theorem~1]{Madiman2008}, and $(d)$ from~\eqref{eq:y2_trans} again.

			\subsubsection{Bounding Lemma~\ref{lem:lemma_bound}}\label{se:proof_lemma_ch5_bound_c}
		
			In the previous step, we have lower-bounded $H\big(\bYbar^{[n]}_{2} |\bH\nt,W_2\big) + H\big(\bigcap_{j>k} \Ybar^{[n]}_{2,j} |  \bH\nt,W_2\big)$. 
			Let us now recover~\eqref{eq:lemma_proof_zef} and focus on its negative terms. 
			It follows that we can repeat~\eqref{eq:bound_dd} for each one of the $H\big(\bigcap_{{j>k}} L^{b[n]}_{\Ybar,j}(\bXbar\nt_{\varnothing})  \mid\bH\nt,W_2\big)$ terms that appear in~\eqref{eq:lemma_proof_zef}, which sums up $M-N_2$ terms. 
			This yields
				\eqmo{\label{eq:proof_er_r0} 
					&(N_2-k) H\big(\bYbar^{[n]}_{2} \mid\bH\nt,W_2\big) + (M-N_2) H\big(\bigcap_{\mathclap{j>k}}\Ybar^{[n]}_{2,j}   \mid\bH\nt,W_2\big) \\
					&\hspace{12ex} \geq (N_2-k-1) H\big(\bYbar^{[n]}_{2} \mid\bH\nt,W_2\big)  + H\big(\bYbar^{[n]}_{2},\  \bL^{[n]}(\bXbar\nt_{\varnothing}) \mid\bH\nt,W_2\big) \\
						&\hspace{12ex}  \ \quad + (M-N_2) H\big(\bigcap_{\mathclap{j>k+1}} L^{b[n]}_{\Ybar,j}(\bXbar\nt_{\varnothing})  \mid\bH\nt,W_2\big) 
				}		
			where  $\bL^{[n]}(\bXbar\nt_{\varnothing})\triangleq \{L_1^{[n]}(\bXbar\nt_{\varnothing}),\cdots,L_{M-N_2}^{[n]}(\bXbar\nt_{\varnothing})\}$ is composed of $M-N_2$ independent linear combinations of~$\bXbar\nt_{\varnothing}$. 
			Now, 	we can further repeat~\eqref{eq:proof_er_r0}  for $j = \{k+1$, $k+2$, $\dots$, $N_2\}$ up to $N_2-k$ times in order to obtain
				\eqmo{\label{eq:proof_er_r} 
					&(N_2-k) H\big(\bYbar^{[n]}_{2} \mid\bH\nt,W_2\big) + (M-N_2) H\big(\bigcap_{\mathclap{j>k}}\Ybar^{[n]}_{2,j}   \mid\bH\nt,W_2\big) \\
					&\hspace{37ex}\geq (N_2-k) 
					H\big(\bYbar^{[n]}_{2},\  \bL^{[n]}(\bXbar\nt_{\varnothing}) \mid\bH\nt,W_2\big). 
				}						
			Note that the entropy terms $H\big(\bYbar^{[n]}_{2},\  \bL^{[n]}(\bXbar\nt_{\varnothing}) \mid\bH\nt,W_2\big)$ are composed of $M$ independent linear combinations of the transmitted signals $\{\Xbar\nt_i\}_{i\in[M]}$, such that it follows that
				\eqm{\label{eq:proof_last_a}
					H\big(\bYbar^{[n]}_{1}\mid \bH\nt, W_2) - H\big(\bYbar^{[n]}_{2},\  \bL^{[n]}(\bXbar\nt_{\varnothing}) \mid\bH\nt,W_2\big)
						& \leq H\big(\bYbar^{[n]}_{1}\mid\bYbar^{[n]}_{2}, \bL^{[n]}(\bXbar\nt_{\varnothing}), \bH\nt, W_2\big)\hspace{-5ex}\nonumber\\
					  & \leq o(n).
				}
			From~\eqref{eq:proof_er_r} and~\eqref{eq:proof_last_a}, it holds that
				\eqmo{
					&(N_2-k) H(\bYbar^{[n]}_{1}\mid\bH\nt,W_2) - (M-k) H(\bYbar^{[n]}_{2}\mid\bH\nt,W_2) \\
					&\hspace{35ex}\leq(N_2-k) H(\bYbar^{[n]}_{1}\mid\bYbar^{[n]}_{2}, \bL^{[n]}(\bXbar\nt_{\varnothing}), \bH\nt, W_2) \\
						&\hspace{35ex}\leq o(n),
				}				
			what concludes the proof of Lemma~\ref{lem:lemma_bound}. \qed

		\section{Achievability Results for the Case $k\geq N_1$}\label{se:dof_achiev}
		The transmission scheme exploits the unavoidable interference as side information, in a similar way as in~\cite{Tandon2012_ISWCS} for the centralized `PD' setting. 
		At the same time, the proposed scheme also exploits the instantaneous CSI available at $\bTX_\star$ by means of the AP-ZF precoding scheme that was introduced in~\cite{Bazco2020_TIT}. 
		The key of the use of AP-ZF is the following lemma (cf.~\cite{Bazco2020_TIT}).
			\begin{lemma}[\!\!\cite{Bazco2020_TIT}]\label{lem:apzf_prop}
				Consider $k$ TXs with perfect CSI and $M-k$ TXs with finite precision CSI. By precoding with AP-ZF the interference can be canceled at $k$ different receive antennas.
			\end{lemma}
		We refer to~\cite{dekerret2012_TIT, Bazco2020_TIT} for more details about AP-ZF. 
			We present in the following the DoF-optimal transmission scheme for  $N_1 \leq k < N_2$, i.e., the proof of  Theorem~\ref{theo:achiev_dof_region}.
			The  achievable scheme for the case $M\leq N_2$ ($\DoF = M$)  is trivial and thus we omit it for sake of conciseness. 
			Given that the DoF does not increase for $M$ bigger than $M = N_1 + N_2$, 
			we consider that  $N_2 < M \leq N_1 + N_2$. 			
						
			We transmit a set $\Sc_i$ of $S_i \triangleq |\Sc_i|$  symbols to RX~$i$, $i\in\{1,2\}$. 
			In particular, we send a total of $S_1 = (M-k)N_1$ symbols to RX~1 and $S_2 = N_2(M-k-N_1) +k N_1$ symbols to RX~2 in a transmission spanning $M-k$ Time Slots (TS). 
 			The scheme is composed of two phases, the first one lasting $N_1$ TS and the second one lasting $M-k-N_1$ TS. 
			Specifically, at each one of the $N_1$ TS of the first phase, we transmit: 
				\itb
					\item $N_1$  independent linear combinations (i.l.c.) of the symbols in $\Sc_1$, which are canceled at $k$ antennas of RX~2 using AP-ZF precoding (see Lemma~\ref{lem:apzf_prop}). 
					\item $M-N_1$ i.l.c. of the symbols in $\Sc_2$, which are canceled at RX~1 through AP-ZF precoding (what is possible  because $k\geq N_1$ and from Lemma~\ref{lem:apzf_prop}).				
				\ite
			Then, at the end of the first phase, 
				\itb
						\item RX~1 has $N_1^2$ i.l.c. of its $S_1 = (M-k)N_1$	symbols. 
						Then, RX~1 needs another $(M-k-N_1)N_1$ i.l.c. to decode all the symbols in $\Sc_1$.  
						\item RX~2 has $N_2 N_1$ i.l.c. of $S_2$ desired	symbols and $(N_2-k)N_1$ interference variables, since the symbols for RX~1 can be canceled only at $k$ of the $N_2$ antennas.  
				\ite
			Let us denote the set of interference terms received at RX~2 during the first phase as $\Ic_2$, $|\Ic_2|=(N_2-k)N_1$. 
			At $\bTX_\star$, we can reconstruct the set $\Ic_2$ thanks to the perfect CSI available. 
			Hence, $\bTX_\star$ can create $(M-k-N_1)N_1$ i.l.c. of $|\Ic_2|$ interference terms, which are functions of the symbols of RX~1, because $|\Ic_2| = M-N_1 \leq N_2$. 
			In the second phase, which lasts $M-k-N_1$ TS, we send at each TS:
				\itb
					\item $N_1$ of the $(M-k-N_1)N_1$ i.l.c. of $\Ic_2$ from $\bTX_\star$.
					\item $N_2-N_1$ i.l.c. of the symbols in $\Sc_2$, which are canceled at RX~1 through AP-ZF precoding. 
				\ite			
			Consequently, at the end of phase~2,			
				\itb
					\item RX~1 has $N_1^2 + (M-k-N_1)N_1 = S_1$ i.l.c. of its $S_1=(M-k)N_1$	symbols.
					Hence, RX~1 can decode all its symbols.				
					\item RX~2 has $N_2 N_1 + N_2(M-k-N_1) = N_2(M-k)$ i.l.c. of $S_2$ desired	symbols and $(N_2-k)N_1$ interference variables, what amounts to 
					$S_2 + (N_2-k)N_1 = N_2(M-k)$ variables. 
					Thus, RX~2 can decode its intended symbols.
				\ite
			Hence, at the end of the communication we have successfully delivered a total of $S_1+S_2 = 
			(M-N_2)N_1 + N_2(M-k)$ symbols over $M-k$ TS, what leads to a sum DoF of
				\eqm{
						d_\Sigma = N_2 + N_1\frac{M-N_2}{M - k},
				}			
			what concludes the proof of Theorem~\ref{theo:achiev_dof_region}. \qed
				
		\section{On the Achievability for the Case $k< N_1$}\label{se:case_m_min}
		In this section, we analyze the achievability results for the case in which $k<N_1$. 
		First, we prove the achievable DoF presented in Proposition~\ref{prop:achiev_m_min_bad}, which serves as lower bound for any configuration. 
		After that, we present a particular case that shows that the lower bound can be improved for certain configurations. 
		
			\subsection{Proof of Proposition~\ref{prop:achiev_m_min_bad}}\label{se:dof_achiev_m_min_bado}
			We present here an achievable scheme attaining a DoF of
				\eqm{
					d_\Sigma
							& = 
									\max\bigg(\min(N_2,M),\ \min(N_2,\; M-k) + \frac{k^2}{\min(N_2,\; M-k)}\ \bigg),				
				}
			for the regime in which $k<N_1$. 
			Let us introduce the notation $\minA \triangleq \min(N_2,\; M-k)$. 
			This scheme is an extension of the scheme presented in Section~\ref{se:dof_achiev}. Therefore, it is composed of two phases of different duration. 
			In this case, we transmit $k+\minA$ symbols per TS during the first $k$~TS. 
			In particular, we transmit: 
				\itb\setlength\itemsep{0pt}
					\item $k$ symbols to RX 2, which are canceled at $k$ antennas of RX 1 by using AP-ZF.
					\item $\minA$ symbols to RX 1, which are canceled at $k$ antennas of RX 2 by using AP-ZF.
				\ite
			Thus, RX~2 can decode its own symbols since it has $k$ antennas free of interference and $k$ symbols to decode.
			RX~2 can then remove the contribution of its own symbols and obtain $\minA-k$ independent linear combinations of the symbols intended by RX~1. 
			Let us denote the set of interference terms received at RX$~2$ as $\Ic_2$, $|\Ic_2|=\minA-k$.
			
			If RX~1 obtains the $\minA-k$ independent linear combinations of its own symbols in  set $\Ic_2$, RX~1 can decode all the $\minA$ symbols, since it has already $k$ linear combinations free of interference. 
			RX~2 already knows those retransmitted symbols, and thus they do not hurt its DoF. 

			In the following $\minA-k$ TS, at each TS we send $\minA$ symbols to RX~2 while retransmitting $k$ of the interference terms in set $\Ic_2$ (the interference received at RX~2 during the first phase).
				\itb\setlength\itemsep{0pt}
						\item The interference retransmitted can be removed perfectly at RX~2, then RX~2 can decode perfectly its own $\minA$ symbols.
						\item The symbols intended by RX~2 are canceled at $k$ antennas of RX~1 thanks to AP-ZF.
						\item RX~1 has $k$ antennas free of interference, and thus it can decode the $k$ retransmitted interference terms.
				\ite
			Consequently, we obtain a DoF of
				\eqm{
						\frac{1}{k+\minA-k}\big(k(k+\minA) + (\minA-k)\minA\big) = \minA + \frac{k^2}{\minA}.
				}

		\subsection{Achievability for the Case $(M,N_1,N_2)=(6,3,3)$}\label{se:m6331}
\begin{figure}[t]\centering%
					
		\centering%
		\begin{tikzpicture}[thick,scale=0.3,>=stealth]
				\draw[] (0,0) node (X1){$X_1$};
				\draw[] (0,5) node (X2){$X_2$};
				\draw[] (0,8) node (X3){$X_3$};
				\draw[] (0,11) node (X4){$X_4$};
				\draw[] (0,14) node (X5){$X_5$};
				\draw[] (0,17) node (X6){$X_6$};

				\draw[] ([xshift=1.5cm, yshift=-1.5cm]X1.south) node (h){\large $~~~\bH$};
				\draw[->, line width =.5pt] ([xshift=1.5cm, yshift=-1.5cm]X1.south) -- ([xshift=-0cm, yshift=-1.5cm]X1.south) -- ([xshift=-0.0cm, yshift=-.7cm]X1.south); 
				
				\draw (-1.5,-1.5) rectangle (1.5,1.5);			
				\draw (-1.5, 3.5) rectangle (1.5,18.5);			
								
				\draw[] (13,0) node (y1){$Y_{1,1}$};
				\draw[] (13,3) node (y2){$Y_{1,2}$};
				\draw[] (13,6) node (y3){$Y_{1,3}$};
				\draw[] (13,11) node (y4){$Y_{2,1}$};
				\draw[] (13,14) node (y6){$Y_{2,2}$};
				\draw[] (13,17) node (y5){$Y_{2,3}$};

				\draw[] (17,3) node (rx1){$\RX 1$};
				\draw[] (17,14) node (rx1){$\RX 2$};
				
				\draw (11.25,-1.5) rectangle (14.5,7.5);			
				\draw (11.25,9.5) rectangle (14.5,18.5);			
							
    \node[myAntennaTX, anchor=south west] at ([xshift=0.3cm]X1.east) {};
    \node[myAntennaTX, anchor=south west] at ([xshift=0.3cm]X2.east) {};
    \node[myAntennaTX, anchor=south west] at ([xshift=0.3cm]X3.east) {};
    \node[myAntennaTX, anchor=south west] at ([xshift=0.3cm]X4.east) {};
    \node[myAntennaTX, anchor=south west] at ([xshift=0.3cm]X5.east) {};
    \node[myAntennaTX, anchor=south west] at ([xshift=0.3cm]X6.east) {};
		
    \node[myAntennaRX, anchor=south east] at ([xshift=-0.3cm]y1.west) {};
    \node[myAntennaRX, anchor=south east] at ([xshift=-0.3cm]y2.west) {};
    \node[myAntennaRX, anchor=south east] at ([xshift=-0.3cm]y3.west) {};
    \node[myAntennaRX, anchor=south east] at ([xshift=-0.3cm]y4.west) {};
    \node[myAntennaRX, anchor=south east] at ([xshift=-0.3cm]y5.west) {};
    \node[myAntennaRX, anchor=south east] at ([xshift=-0.3cm]y6.west) {};
			
		\draw [dotted, line width=.5pt,  decorate,decoration={calligraphic brace,amplitude=8pt},xshift=-0pt,yshift=0pt] (-2.5,3.5) -- (-2.5,18.5) node [black,midway,xshift=-5.5ex] {%
				\begin{tabular}{c} ~\\[.75ex] ~~~$\bTX_\varnothing$  \\[.85ex] \footnotesize $(M-k = 5)$~~~ \end{tabular}%
			};
		\draw [dotted, line width=0.5pt, decorate,decoration={calligraphic brace,amplitude=7pt},xshift=-0pt,yshift=0pt] (-2.5,-2.25) -- (-2.5,2.25) node [black,midway,xshift=-5.5ex] {%
				\begin{tabular}{c}~\\[0.5ex] ~~~$\bTX_\star$  \\[.5ex] \footnotesize $~(k = 1)$ \end{tabular}%
			};
			
	\end{tikzpicture}~\vspace{-2ex}		
				\caption{Equivalent channel for the case $(M,N_1,N_2,k)=(6,3,3,1)$.}\label{fig:equiv_channel_1}
			\end{figure}
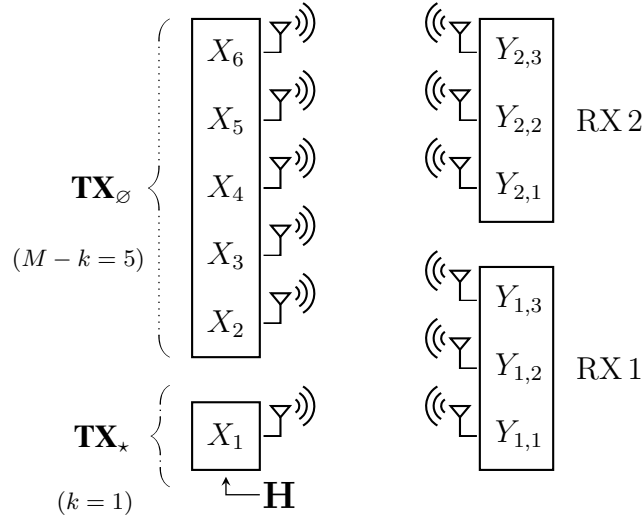

		Let us consider a setting with $M=6$ transmit antennas and $N_1 = N_2 = 3$ antennas at each RX. 
		Suppose that only one transmit antenna has perfect CSI for the whole channel matrix, while the other 5~transmit antennas have only finite precision CSI. 
		Thus, $k=1$. 
		This setting, denoted as $(M,N_1,N_2,k) = (6,3,3,1)$, is illustrated in Fig.~\ref{fig:equiv_channel_1}. 
		We present here a scheme that achieves a sum DoF of 4.

		The scheme consists of two phases, each one of 2 Time Slots (TS), and it is presented in Table~\ref{tab:achiev_6331}, in which every row represents one antenna. 
		The table is divided in three horizontal parts: 
		the top part represents the symbols transmitted from each antenna, the middle part represents the received signal at RX~1, and the bottom part shows the received signal at TX~2. 
			\begin{table}[t]\centering\small
				\caption[Achievability for the~$(M,N_1,N_2,k)=(6,3,3,1)$ setting]{Description of a transmission scheme achieving the optimal $\DoF=4$ for the setting~$(M,N_1,N_2,k)=(6,3,3,1)$.} \label{tab:achiev_6331}
				\begin{tabular}{ccccc}
						&	$t=1$	&	$t=2$	&	$t=3$	&	$t=4$ \\ \hline
						\phantom{e}  & & & & \\[-2ex]
						$\bTX_{\star,1} $		& $c+ f(a_1,a_2,a_3,a_4,a_5)$ & $c+ f'(b_1,b_2,b_3,b_4,b_5)$ & $d\!+\!f''(a_6,a_7,a_8)$ & $d\!+\! f'''(b_6,b_7,b_8)$ \\ 
						$\bTX_{\varnothing,1}\!\!	$	& $a_1,a_2,a_3,a_4,a_5$ & $b_1,b_2,b_3,b_4,b_5$ & $a_6,a_7,a_8$ & $b_6,b_7,b_8$\\ 
						$\bTX_{\varnothing,2}\!\!	$	& $a_1,a_2,a_3,a_4,a_5$ & $b_1,b_2,b_3,b_4,b_5$ & $a_6,a_7,a_8$&$b_6,b_7,b_8$ \\ 
						$\bTX_{\varnothing,3}\!\! $	& $a_1,a_2,a_3,a_4,a_5$ & $b_1,b_2,b_3,b_4,b_5$ & $a_6,a_7,a_8$&$b_6,b_7,b_8$ \\ 
						$\bTX_{\varnothing,4}\!\!	$	& $a_1,a_2,a_3,a_4,a_5$ & $b_1,b_2,b_3,b_4,b_5$ & $a_6,a_7,a_8$&$b_6,b_7,b_8$ \\ 
						$\bTX_{\varnothing,5}\!\!	$	& $a_1,a_2,a_3,a_4,a_5$ & $b_1,b_2,b_3,b_4,b_5$ & $a_6,a_7,a_8$&$b_6,b_7,b_8$ \\
						\phantom{e}  & & & & \\[-2ex]
						\hline
						\phantom{e}  & & & & \\[-2ex]
						$Y_{1,1}$		& $f^1_1(a_1,a_2,a_3,a_4,a_5,c)$ & $f^1_4(b_1,b_2,b_3,b_4,b_5,c)$ 
																& $g^1_1(a_6,a_7,a_8,d)$ & $g^1_4(b_6,b_7,b_8,d)$ \\ 
						$Y_{1,2}$		& $f^1_2(a_1,a_2,a_3,a_4,a_5,c)$ & $f^1_5(b_1,b_2,b_3,b_4,b_5,c)$ 
																& $g^1_2(a_6,a_7,a_8,d)$ & $g^1_5(b_6,b_7,b_8,d)$ \\ 
						$Y_{1,3}$   & $f^1_3(a_1,a_2,a_3,a_4,a_5,c)$ & $c$ 
																& $g^1_3(a_6,a_7,a_8,d)$ & $d$ \\ 
						\phantom{e}  & & & & \\[-2ex]
						\hline		
						\phantom{e}  & & & & \\[-2ex]
						$Y_{2,1}$	& $f^2_1(a_1,a_2,a_3,a_4,a_5,c)$ & $f^2_3(b_1,b_2,b_3,b_4,b_5,c)$  
																& $g^2_1(a_6,a_7,a_8,d)$ & $g^2_3(b_6,b_7,b_8,d)$ \\  
						$Y_{2,2}$	& $f^2_2(a_1,a_2,a_3,a_4,a_5,c)$ & $f^2_4(b_1,b_2,b_3,b_4,b_5,c)$  
																& $g^2_2(a_6,a_7,a_8,d)$ & $g^2_4(b_6,b_7,b_8,d)$ \\
						$Y_{2,3}$	& $c$ & $f^2_5(b_1,b_2,b_3,b_4,b_5,c)$  
																& $d$ & $g^2_5(b_6,b_7,b_8,d)$ \\
						\phantom{e}  & & & & \\[-2ex]
						\hline
				\end{tabular}%
			\end{table}%
		%
		Let us disclose the previous table by describing the transmission scheme. 
		
		\paragraph{Encoding}
		We send 18~symbols ($a_{1-8},b_{1-8},c,d$). 
		Symbols $a_i$ are intended to RX~1 and symbols $b_i$ are intended to RX~2. 
		The functions $f$, $f'$, $f''$, and $f'''$ are such that the corresponding symbols $a_i$ or $b_i$ are canceled at the third antenna of the non-intended RX. 
		The functions $f^i_j$ and $g^i_j$ are defined such that they represent the received signal at RX~$i$. 
		The sub-index~$j$ is used to order and identify the different received signals. 
		$f^i_j$ denotes the received signal during the first two TS, whereas $g^i_j$ denotes the received signal for the last two TS. 
		
		\paragraph{Decoding}
		First, the symbols $c$ and $d$ can be decoded at both RXs from the received signal at their third antenna: RX$~1$ decodes $c$ at $t=2$ and $d$ at $t=4$, while RX$~2$ do so at $t=1$ and $t=3$, respectively.  
		Then, after subtracting $d$ from the received signals, the messages $a_{6-8}$ and  $b_{6-8}$ are easily obtained at the intended RX from the received signal of $t=3$ and $t=4$. 
		Furthermore,  if RX~1 obtained equations $f^2_1$ and $f^2_2$, it would be able to decode all the $a_{1-5}$. 
		Similarly, if RX~2 obtained equations $f^1_4$ and $f^1_5$, it would be able to decode all the $b_{1-5}$. 
		Hence, we select $c$ and $d$ as
		\begin{align}
				c &= f^2_1 \oplus f^1_4, \\
				d &= f^2_2 \oplus f^1_5. 
		\end{align}
		Therefore, RX~1 can subtract $f^1_4$ from $c$  and $f^1_5$ from $d$ and obtain the necessary equations. 
		On the other hand, RX~2 can subtract $f^2_1$ from $c$  and $f^2_2$ from $d$ and  obtain also the necessary equations.
		Since we have causal CSIT, we can not encode the signal $f^1_4$ of $t=2$ in $c$, but we can accept a one-block delay and transmit the received signal of the previous transmission block. 
		The DoF loss  will be negligible if the time considered is long enough. 
		For $t=3$ and $t=4$, RX~1 obtains $d$ in $t=4$ and, after that, it can decode $a_6,a_7,a_8$ from $t=3$. 
		In the same way, RX~2 obtains $d$ in $t=3$ and thus it can decode $b_6,b_7,b_8$ at $t=4$.
		
		Consequently, we transmit 16~information symbols in 4~TS, and thus $\DoF=4$. 
		The general achievable scheme presented in Section~\ref{se:dof_achiev_m_min_bado} only attains a DoF of~$\frac{10}{3}$, whereas the upper bound of Lemma~\ref{lem:s_dof_big} yields $\DoF\leq 4 + \frac{4}{5}$.
		Interestingly, the sum DoF of the the $(M,N_1,N_2) = (6,3,3)$ setting is bounded by
			\begin{table*}[h]\centering%
				\begin{tabular}{llll}\normalsize
						$\DoF = 3$ &\quad \normalsize if $k = 0$, \hspace{10ex} 
									&\normalsize $4 + \frac{1}{3} \leq \DoF \leq 5+\frac{1}{4}$  &\quad \normalsize if $k = 2$, \\ \normalsize
						$4 \leq \DoF \leq	4 + \frac{4}{5}$ &\quad \normalsize if  $k = 1$, 
									& \normalsize $\DoF =	6$ &\quad \normalsize if  $k \geq 3$.
				\end{tabular}%
			\end{table*}%
			
		From previous insights obtained for other settings\cite{Bazco2018_WCL,Bazco2020_TIT} and the intuition that one extra informed antenna cannot bring out more than one DoF, we could conjecture that  $\DoF = 4$ if $k=1$ and $\DoF = 5$ if $k=2$. 
		However, this characterization remains an open and interesting problem. 
		
	
		\section{Conclusion}\label{se:conclusion_ch5_upper}
		We have analyzed the 2-user MIMO BC setting in which only $k$ transmit antennas have access to perfect CSI, whereas the other $M-k$ transmit antennas have access only to finite precision CSI. 
		We have derived an outer bound for the DoF region that is tight for $k\geq\min(N_1,N_2)$, characterizing the loss of DoF obtained from reducing the number of informed antennas.  
		On this basis, we have shown that it is not necessary to have perfect CSI at every transmit antenna, but only at~$\max(N_1,N_2)$ antennas. 
		We have also presented an achievable scheme that adapts to the distributed CSI setting so as to boost the DoF with respect to the use of conventional centralized schemes.

		\begin{appendix}	
											
			\section*{Converse of Theorem~\ref{theo:dof_r_big} for the case $M > N_1+N_2$}\label{se:region_outerbound_b}
			We define $N \triangleq N_1+N_2$. 
			We recall that the $M$ transmit antennas are divided as
				\eqm{
					\bTX	\triangleq \big[
								\underbrace{\TX_1,\ \dots,\ \TX_k}_{\bTX_{\star}},\  
								\underbrace{\TX_{k+1},\ \dots,\ \TX_{M}}_{\bTX_{\varnothing}}\big]				
				}
			The channel $\bH \in \Cb^{N\times M}$ has $M-N$ null space dimensions. 
			Therefore, if we could apply a rotation matrix $\bR$ with unit determinant to  make $\bH\bR$'s right $M-N$ columns be zero, it would lead to an equivalent channel where the RXs do not listen to the last $(M-N)$ TXs. Consider $\bH' \triangleq \bH\bR$. Then, 
				\eqm{
					\bH' = \left[\begin{array}{cc}
											\bH'_{[1:N,1:N]} & \mathbf{0}_{N\times (M-N)}
									\end{array}\right].
				} 
			In order to obtain this equivalent channel, we apply an invertible linear transformation at the transmit antennas by multiplying the transmit signal $\bX$ by $\bR$. 
			Hence, we transmit $\bX' \triangleq \bR\bX$ in place of $\bX$. 
			After this transformation, we can derive the upper bound by applying the same steps as in Section \ref{se:region_outerbound_a} for $M = N_1 + N_2$, since the RXs only listen to $N_1 + N_2$ transmit antennas. 

			\subsubsection{Channel Rotation with Distributed CSIT}

				Although it is straightforward to apply the previous channel transformation in a centralized scenario where all the transmit antennas are seen as one single entity, it is not direct that it can be applied in our distributed scenario, where every single transmit antenna is isolated with respect to the others and has to act \emph{only} based on his own local information. 
				Thereupon, we show that this channel transformation is possible in the decentralized MIMO BC.

				In the $(M,N_1,N_2,k)$ scenario considered, the matrix multiplication $\bR\bX$ must be done \emph{locally}. 
				Consequently, the equivalent transmitted signal at $\TX_i$, $X_i$, is obtained as
				\eqm{
						X'_i = \bR_i \bX,
				}
				where $\bR_i$ is the $i$-th row of $\bR$. 
				However, the $M-k$ antennas with finite precision CSIT ($\bTX_\varnothing$) are not able to obtain neither $\bR$ nor the transmit signal from the TXs with perfect CSIT ($\bTX_\star$).  
				In order to deal with this problem, we first let all the TXs in $\bTX_\star$ cooperate among them. 
				Similarly, we let the TXs in $\bTX_\varnothing$ cooperate among them. 
				Since every TX in $\bTX_\star$ already had perfect information of the whole channel, assuming that they are a unique transmitter with $k$ antennas does not affect the analysis. 
				In the same way, assuming that the $M-k$ TXs with finite precision CSI form a unique transmitter with $M-k$ antennas does not give any improvement to them, since they still have only finite precision CSI. 
				Furthermore, cooperation can not hurt. 
				Therefore, we have an equivalent channel with two TXs, $\bTX_\star$ that transmits $\bX_\star$, and $\bTX_\varnothing$ that transmits $\bX_\varnothing$.  
				The channel transformation is applied as
					\eqm{
						\bX'_\star &= \bR_{[1:k,1:M]}\bX\\
						\bX'_\varnothing &= \bR_{[k+1:M,1:M]}\bX
					}

			\subsubsection{Composition of the Transformation Matrix}
				We aim to obtain a matrix $\bR \in \Cb^{N\times M}$ such that $\bH' \triangleq \bH\bR$ satisfies 
					\eqm{
						\bH' = \left[\begin{array}{cc}
											\bH'_{[1:N,1:N]} & \mathbf{0}_{N\times (M-N)} \label{eq:gprime_def}
									\end{array}\right].
					} 
				In order to obtain \eqref{eq:gprime_def}, we need $h'_{i,j} = 0$, for any $j \in \{N+1,\dots,M\}$ and any~$i$. 
				In order to transform the $j$-th channel column, we solve the following linear system
					\eqm{
						\left[\begin{array}{cccc}
									h_{1,1} & h_{1,2} & \dots & h_{1,N}\\
									\vdots & \vdots & \ddots & \vdots\\
									h_{N,1} & h_{N,2} & \dots & h_{N,N}
						\end{array}\right]
						\left[\begin{array}{c}
									r_{1,j}\\
									\vdots\\
									r_{N,j}
						\end{array}\right] = 
						\left[\begin{array}{c}
									-h_{1,j}\\
									\vdots\\
									-h_{N,j}
						\end{array}\right].	\label{eq:r_linear_system} 
					}
				From the channel independence assumption,  $\bH_{[1:N,1:N]} $ is full rank almost surely, and therefore the system has a solution. 
				Hence, the matrix $\bR$ is defined as
					\eqm{
						\bR \triangleq
								\left[\begin{array}{cc}
											\bI_{N \times N} 					  & \begin{array}{ccc}
																													r_{1,N+1}	& \dots 	& r_{1,M}\\
																													\vdots			& \ddots	& \dots\\ 
																													r_{N,N+1}	& \dots 	& r_{N,M}
																										\end{array}\\
											\mathbf{0}_{(M-N) \times N}	& \bI_{(M-N) \times (M-N)}
								\end{array}\right].		 \label{eq:matrix_rot_def}
					}
				From \eqref{eq:matrix_rot_def}, it holds that $\bH'_{[1:N,1:N]} = \bH_{[1:N,1:N]}$. 
				Note that the antennas with finite precision CSI can obtain their equivalent transmit signals as
					\eqmo{
						\bX'_\varnothing &= \bR_{[k+1:M,1:M]} \bX\\
									&=  \underbrace{\left[\begin{array}{ccccccc}
												\bR_{k+1,k+1}	&	\dots		&	\bR_{k+1,M}\\
												\vdots					&	\ddots	&	\vdots\\
												\bR_{M,k+1}			&	\dots	  &	\bR_{M,M}
									\end{array}\right]}_{\bR_\varnothing} \bX_\varnothing.
					}
				Therefore, the transformation at the TXs with finite precision depends only on their own transmit signals and they do not need to know $\bX_\star$. 
				Furthermore, let us consider that a genie provides $\bTX_\varnothing$ with the matrix $\bR_\varnothing$. In this case,~\eqref{eq:r_linear_system} and the finite precision CSIT assumption imply that $\bTX_\varnothing$ can not infer any $h_{i,j}$ from the knowledge of $\bR_\varnothing$. 
				Hence, we can apply the transformation in the Distributed MIMO BC setting, which concludes the proof of  Theorem~\ref{theo:dof_r_big}. 

		\end{appendix}	

			\bibliographystyle{IEEEtran}												
			\bibliography{IEEEabrv,./Literature}													

\begin{thebibliography}{10}
\providecommand{\url}[1]{#1}
\csname url@samestyle\endcsname
\providecommand{\newblock}{\relax}
\providecommand{\bibinfo}[2]{#2}
\providecommand{\BIBentrySTDinterwordspacing}{\spaceskip=0pt\relax}
\providecommand{\BIBentryALTinterwordstretchfactor}{4}
\providecommand{\BIBentryALTinterwordspacing}{\spaceskip=\fontdimen2\font plus
\BIBentryALTinterwordstretchfactor\fontdimen3\font minus
  \fontdimen4\font\relax}
\providecommand{\BIBforeignlanguage}[2]{{%
\expandafter\ifx\csname l@#1\endcsname\relax
\typeout{** WARNING: IEEEtran.bst: No hyphenation pattern has been}%
\typeout{** loaded for the language `#1'. Using the pattern for}%
\typeout{** the default language instead.}%
\else
\language=\csname l@#1\endcsname
\fi
#2}}
\providecommand{\BIBdecl}{\relax}
\BIBdecl

\bibitem{Davoodi2016_TIT_DoF}
{A. G. Davoodi and S. A. Jafar}, ``Aligned image sets under channel
  uncertainty: Settling conjectures on the collapse of {Degrees of Freedom}
  under finite precision {CSIT},'' \emph{{IEEE} Trans. Inf. Theory}, vol.~62,
  no.~10, pp. 5603--5618, Oct 2016.

\bibitem{MaddahAli2012}
M.~A. {Maddah-Ali} and D.~{Tse}, ``Completely stale transmitter channel state
  information is still very useful,'' \emph{{IEEE} Trans. Inf. Theory},
  vol.~58, no.~7, pp. 4418--4431, July 2012.

\bibitem{Davoodi2019_ISIT}
A.~G. {Davoodi} and S.~A. {Jafar}, ``{Degrees of Freedom} region of the {(M,
  N1, N2) MIMO} broadcast channel with partial {CSIT}: An application of
  sum-set inequalities,'' in \emph{{Proc. IEEE International Symposium on
  Information Theory (ISIT)}}, July 2019, pp. 1637--1641.

\bibitem{Wang2017_TIT}
Y.~{Wang} and M.~K. {Varanasi}, ``{Degrees of Freedom} of the two-user {MIMO}
  broadcast channel with private and common messages under hybrid {CSIT}
  models,'' \emph{{IEEE} Trans. Inf. Theory}, vol.~63, no.~9, pp. 6004--6019,
  Sep. 2017.

\bibitem{Tandon2012b}
R.~{Tandon}, S.~A. {Jafar}, S.~{Shamai}, and H.~V. {Poor}, ``On the synergistic
  benefits of alternating {CSIT} for the {MISO} broadcast channel,''
  \emph{{IEEE} Trans. Inf. Theory}, vol.~59, no.~7, pp. 4106--4128, July 2013.

\bibitem{Rassouli2016}
B.~Rassouli, C.~Hao, and B.~Clerckx, ``{DoF} analysis of the {MIMO} broadcast
  channel with alternating/hybrid {CSIT},'' \emph{{IEEE} Trans. Inf. Theory},
  vol.~62, no.~3, pp. 1312--1325, Mar. 2016.

\bibitem{Aggarwal2012}
V.~{Aggarwal}, Y.~{Liu}, and A.~{Sabharwal}, ``Sum capacity of interference
  channels with a local view: Impact of distributed decisions,'' \emph{{IEEE}
  Trans. Inf. Theory}, vol.~58, no.~3, pp. 1630--1659, March 2012.

\bibitem{Bazco2018_WCL}
A.~Bazco-Nogueras, P.~de~Kerret, D.~Gesbert, and N.~Gresset, ``{Distributed
  CSIT} does not reduce the {Generalized DoF} of the 2-user {MISO} broadcast
  channel,'' \emph{{IEEE} Wireless Commun. Lett.}, vol.~8, no.~3, pp. 685--688,
  June 2019.

\bibitem{Bazco2020_TIT}
A.~{Bazco-Nogueras}, P.~{de Kerret}, D.~{Gesbert}, and N.~{Gresset}, ``On the
  {Degrees-of-Freedom} of the {K}-user distributed broadcast channel,''
  \emph{{IEEE} Trans. Inf. Theory}, pp. 1--1, 2020, early access.

\bibitem{Tandon2012_ISWCS}
R.~{Tandon}, M.~A. {Maddah-Ali}, A.~{Tulino}, H.~V. {Poor}, and S.~{Shamai},
  ``On fading broadcast channels with partial channel state information at the
  transmitter,'' in \emph{{Int. Symp. on Wireless Commun. Syst. (ISWCS)}}, Aug
  2012, pp. 1004--1008.

\bibitem{ElGamal2011}
A.~El~Gamal and Y.~Han~Kim, \emph{{Network information theory}}.\hskip 1em plus
  0.5em minus 0.4em\relax Cambridge University Press, 2011.

\bibitem{Davoodi2017_TIT_GDoF_IC_finite}
A.~G. {Davoodi} and S.~A. {Jafar}, ``Generalized {Degrees of Freedom} of the
  symmetric {$K$} user interference channel under finite precision {CSIT},''
  \emph{{IEEE} Trans. Inf. Theory}, vol.~63, no.~10, pp. 6561--6572, Oct 2017.

\bibitem{Davoodi2017_ISIT_sumset}
A.~G. Davoodi and S.~A. Jafar, ``{Sum-set inequalities from aligned image sets:
  Instruments for robust GDoF bounds},'' in \emph{{Proc. IEEE International
  Symposium on Information Theory (ISIT)}}, June 2017, pp. 684--688.

\bibitem{Hao2017}
C.~{Hao}, B.~{Rassouli}, and B.~{Clerckx}, ``Achievable {DoF} regions of {MIMO}
  networks with imperfect {CSIT},'' \emph{{IEEE} Trans. Inf. Theory}, vol.~63,
  no.~10, pp. 6587--6606, Oct 2017.

\bibitem{Bresler2008}
\BIBentryALTinterwordspacing
G.~Bresler and D.~Tse, ``{The two-user Gaussian interference channel: A
  deterministic view},'' \emph{{European Transactions on Telecommunications}},
  vol.~19, no.~4, pp. 333--354, 2008. [Online]. Available:
  \url{https://onlinelibrary.wiley.com/doi/abs/10.1002/ett.1287}
\BIBentrySTDinterwordspacing

\bibitem{Madiman2008}
M.~{Madiman}, ``On the entropy of sums,'' in \emph{Proc. IEEE Information
  Theory Workshop (ITW)}, May 2008, pp. 303--307.

\bibitem{dekerret2012_TIT}
P.~de~Kerret and D.~Gesbert, ``{Degrees of freedom of the network MIMO channel
  with distributed CSI},'' \emph{{IEEE} Trans. Inf. Theory}, vol.~58, no.~11,
  pp. 6806--6824, Nov. 2012.

\end{thebibliography}

\end{document}